\documentclass[12pt,a4paper]{article}

\usepackage{amsmath,amssymb,amsthm}
\usepackage{graphicx}
\usepackage[dvipsnames]{xcolor}
\usepackage{hyperref}
\usepackage{geometry}
\usepackage{booktabs}
\usepackage{tikz}
\usetikzlibrary{arrows.meta, positioning, shapes.geometric,
                decorations.pathreplacing, calc, fit, backgrounds}
\usepackage{cite}
\usepackage{authblk}
\usepackage{microtype}
\usepackage{enumitem}
\usepackage{caption}
\usepackage{float}
\usepackage{array}
\usepackage{bm}

\hypersetup{
  colorlinks=true, linkcolor=NavyBlue, citecolor=NavyBlue, urlcolor=NavyBlue,
  pdftitle={Knowledge Manifold: A Riemannian Geometric Framework for Semantic Mapping and Geodesic Analysis of Scientific Literature},
  pdfauthor={Tomonaga Okabe, Kazuhiko Komatsu}
}

\theoremstyle{definition}

\theoremstyle{plain}

\theoremstyle{remark}

\newcommand{\vv}[1]{\bm{#1}}
\newcommand{\norm}[1]{\left\|#1\right\|}
\newcommand{\T}{^{\mathsf{T}}}
\newcommand{\pd}[2]{\dfrac{\partial #1}{\partial #2}}

\title{\textbf{Knowledge Manifold: A Riemannian Geometric Framework for\\
       Semantic Mapping and Geodesic Analysis of Scientific Literature}}

\author[1,2]{Tomonaga Okabe\thanks{E-mail:
  \href{mailto:tomonaga.okabe.a8@tohoku.ac.jp}{tomonaga.okabe.a8@tohoku.ac.jp}}}
\author[2]{Kazuhiko Komatsu\thanks{Corresponding author.
  E-mail: \href{mailto:komatsu@tohoku.ac.jp}{komatsu@tohoku.ac.jp}}}
\affil[1]{Department of Aerospace Engineering, Tohoku University,
          6-6-01, Aoba, Aramaki, Aoba-ku, Sendai, Miyagi 980-8579, Japan}
\affil[2]{Research Center for Green X-Tech, Tohoku University,
          6-6-11, Aoba, Aramaki, Aoba-ku, Sendai, Miyagi 980-8579, Japan}
\date{June 3, 2026}

\begin{document}
\maketitle

\begin{abstract}
We present the \emph{knowledge manifold}: a Riemannian geometric space in
which a corpus of documents is arranged according to semantic positional
relationships derived from character $n$-gram TF-IDF representations.
The framework proceeds in five tightly coupled stages.
First, each document is converted to a character-level $n$-gram TF-IDF
vector (4-7 grams, up to 250,000 features, $\ell^2$-normalized) and
embedded in a two-dimensional knowledge map via constrained stress
minimization with repulsion, variance, and centering regularizers.
Second, knowledge at an arbitrary query point is estimated through
Smoothed Particle Hydrodynamics (SPH) interpolation using a cubic-spline
kernel, yielding an interpolated TF-IDF feature vector that can be
linguistically characterized.
Third, directional knowledge gradients at $0^\circ$, $45^\circ$, and
$90^\circ$ are computed from the SPH interpolation map, and pairwise
directional similarity is quantified via inner product and cosine
similarity.
Fourth, a Gaussian Process Regression (GPR) model, with a
Constant$\times$RBF$+$White kernel fitted on a 10-dimensional SVD
projection, provides a Bayesian posterior mean, uncertainty estimate,
and per-document contribution rate at the query point.
Fifth, geodesics in the knowledge space are obtained by minimizing a
discrete Riemannian path energy derived from the SPH-induced metric
tensor, using $L$-BFGS-B with seven deterministic initial-path
candidates.
We apply the formulation to a corpus of 20 papers in fiber-reinforced
composite materials and aerospace structural mechanics, showing that the
semantic map recovers meaningful research clusters, geodesic paths reveal
natural conceptual bridges between distant topics, and SPH/GPR
interpolation enables the generation of \emph{virtual knowledge}: hypothetical paper abstracts describing unstudied but geometrically
predicted research directions.

\end{abstract}

\section{Introduction}

The exponential growth of scientific literature demands tools that go
beyond keyword search or citation analysis.  A complementary perspective
is to treat the body of literature as a continuous geometric
object, a \emph{knowledge manifold}, where each document occupies a
position determined by its semantic content, and distances encode
conceptual dissimilarity.

Our framework has three distinguishing characteristics.
\begin{enumerate}[label=(\roman*)]
  \item \textbf{Character $n$-gram TF-IDF.}
    Unlike word-level or neural-embedding approaches, character $n$-gram
    features are robust to domain-specific compound words, abbreviations,
    and translingual technical notation, which are prevalent in materials
    science and aerospace engineering.
  \item \textbf{Physics-inspired interpolation (SPH).}
    Smoothed Particle Hydrodynamics~\cite{Gingold1977} treats each
    document as a ``particle'' and estimates the knowledge density at
    arbitrary query points via a cubic-spline kernel.  This provides
    both an interpolated knowledge vector and its spatial gradients,
    collectively termed the \emph{knowledge gradient}, which reveal the directions of
    fastest semantic change.
  \item \textbf{Geodesic analysis.}
    Geodesics on the knowledge manifold are the intrinsically shortest
    paths between two research areas, accounting for the curvature of
    the semantic space induced by the SPH map.  Unlike Euclidean
    straight lines in the projected 2-D space, geodesics traverse regions
    of high document density and hence represent the most natural
    conceptual transition routes.
\end{enumerate}

The main contributions of this paper are:
\begin{enumerate}[label=(\roman*)]
  \item A complete, reproducible specification of the knowledge manifold
        formulation with all hyper-parameters fixed
        (Section~\ref{sec:formulation}).
  \item A mathematically precise formulation of the SPH knowledge
        interpolation and gradient
        (Section~\ref{sec:sph}).
  \item A Bayesian uncertainty quantification scheme via GPR that
        provides per-document contribution rates at any query point
        (Section~\ref{sec:gpr}).
  \item A guaranteed-convergent geodesic solver with deterministic
        multi-start and energy-based success criteria
        (Section~\ref{sec:geodesic}).
  \item Experimental results on a composite-materials corpus
        (Section~\ref{sec:experiment}).
\end{enumerate}

\section{Related Work}

\paragraph{Document representation.}
Bag-of-words models, including TF-IDF~\cite{Sparck1972}, have long been
used for information retrieval.  Character $n$-gram models, introduced
for text categorization~\cite{Cavnar1994}, were later applied in neural
contexts~\cite{Bojanowski2017} but remain highly competitive for
specialized technical corpora.  Dense embedding models, including word2vec
\cite{Mikolov2013}, GloVe~\cite{Pennington2014}, and Sentence-BERT
\cite{Reimers2019}, offer richer semantic information but sacrifice
interpretability.

\paragraph{Manifold learning.}
ISOMAP~\cite{Tenenbaum2000}, UMAP~\cite{McInnes2018}, and
t-SNE~\cite{Maaten2008} embed high-dimensional point clouds in low
dimensions while preserving local or global geometry.  SMACOF/MDS
minimizes a stress criterion \cite{Borg2005} and provides a natural
initialization for our constrained optimization.
In contrast to these approaches, which are designed primarily for
low-dimensional data visualization, the present framework treats the
document corpus as a continuous knowledge manifold and extends the
analysis to geodesic path search driven by a Riemannian metric induced
from SPH interpolation.

\paragraph{SPH in non-physical domains.}
SPH has been applied to image processing~\cite{Mclaughlin1997} and
scattered-data interpolation~\cite{Fasshauer2007}.  Its application to
document-space interpolation is, to our knowledge, novel.

\paragraph{Gaussian Process Regression on text.}
GPR has been used for active learning of document labels \cite{Kapoor2007}
and text regression \cite{Cohn1996}, but rarely for knowledge uncertainty
quantification in literature analysis.

\paragraph{Geodesics for knowledge navigation.}
Shortest-path methods on $k$-NN graphs~\cite{Tenenbaum2000} are standard
in manifold learning; however, geodesics on a \emph{continuously
interpolated} manifold derived from SPH, rather than on a discrete
graph, offer smoother trajectories and explicit curvature information.

\section{Knowledge Manifold Formulation}
\label{sec:formulation}

\subsection{Document Vectorization}

Let $\mathcal{D}=\{d_1,\dots,d_N\}$ ($N=20$) be the document corpus.
Text is extracted in page order from each PDF; preprocessing is limited
to Unicode NFKC normalization, ASCII lowercasing, and collapse of
consecutive whitespace.  No stop-word removal, stemming, or translation
is applied.

Each document $d_i$ is mapped to a TF-IDF feature vector
$\vv{x}_i\in\mathbb{R}^V$ using the following fixed configuration
(scikit-learn \texttt{TfidfVectorizer} equivalent):
\begin{itemize}[noitemsep]
  \item \texttt{analyzer="char\_wb"} (within-word character boundary)
  \item \texttt{ngram\_range=(4,7)} (4- to 7-character grams)
  \item \texttt{sublinear\_tf=True}\ \ (i.e.\ $\mathrm{tf}\leftarrow 1+\log\,\mathrm{tf}$)
  \item \texttt{norm="l2"},\ \texttt{max\_features=250\,000},\ \texttt{min\_df=1}
\end{itemize}
After vectorization each document vector is $\ell^2$-normalized:
$\hat{\vv{x}}_i = \vv{x}_i/\norm{\vv{x}_i}$.
All subsequent similarities are cosine similarities; distances are
cosine distances:
\begin{equation}
  d_{ij} = 1 - \hat{\vv{x}}_i\T\hat{\vv{x}}_j \in [0,\,2].
  \label{eq:cosine_distance}
\end{equation}

\subsection{Selection of Four Representative Documents}
\label{sec:rep}

To anchor the two-dimensional map, four representative documents
$\{r_1,r_2,r_3,r_4\}\subset\mathcal{D}$ are chosen by exhaustive search
to maximize the sum of pairwise cosine distances:
\begin{equation}
  \{r_1,r_2,r_3,r_4\} =
  \arg\max_{\{a,b,c,d\}\subset\mathcal{D}}
  \sum_{(i,j)\in\binom{\{a,b,c,d\}}{2}} d_{ij}.
  \label{eq:rep_selection}
\end{equation}
Ties are broken by lexicographic filename order.  The four representatives
are fixed to the corners of the unit square in lexicographic order:
\begin{equation}
  \vv{y}_{r_1}=(-1,-1),\quad
  \vv{y}_{r_2}=(1,-1),\quad
  \vv{y}_{r_3}=(1,1),\quad
  \vv{y}_{r_4}=(-1,1).
  \label{eq:corners}
\end{equation}

\subsection{Two-Dimensional Knowledge Map}
\label{sec:2dmap}

The two-dimensional coordinates
$\{\vv{y}_i\in\mathbb{R}^2\}_{i=1}^{N}$ for the remaining 16 documents
are obtained by minimizing the objective function
\begin{equation}
  J = E_{\text{stress}}
    + \lambda_{\text{rep}}\,E_{\text{rep}}
    + \lambda_{\text{var}}\,E_{\text{var}}
    + \lambda_{\text{ctr}}\,E_{\text{ctr}},
  \label{eq:objective}
\end{equation}
with $\lambda_{\text{rep}}=0.10$, $\lambda_{\text{var}}=0.05$,
$\lambda_{\text{ctr}}=0.01$.
The stress term follows the standard MDS formulation~\cite{Borg2005},
while the repulsion, variance, and centering terms are introduced as
regularizers to obtain a stable and well-spread knowledge map.

\paragraph{Distance-preservation (stress) term.}
\begin{equation}
  E_{\text{stress}} = \frac{
    \displaystyle\sum_{i<j}
    \Bigl(\norm{\vv{y}_i-\vv{y}_j} - \alpha\,d_{ij}\Bigr)^2
  }{
    \displaystyle\sum_{i<j}\!\bigl(\alpha\,d_{ij}\bigr)^2
  },
  \label{eq:stress}
\end{equation}
where $\alpha$ is a fixed scale factor chosen so that the mean 2-D
distance between the four corner documents equals the mean cosine
distance among those four documents.

\paragraph{Repulsion term.}
\begin{equation}
  E_{\text{rep}} = \frac{2}{N(N-1)}
  \sum_{i<j}\exp\!\left(-\frac{\norm{\vv{y}_i-\vv{y}_j}^2}{2\sigma^2}\right),
  \quad\sigma=0.35.
  \label{eq:repulsion}
\end{equation}

\paragraph{Variance term.}
\begin{equation}
  E_{\text{var}} = \frac{(\mathrm{var}_x - \tfrac{1}{3})^2
                        +(\mathrm{var}_y - \tfrac{1}{3})^2}
                       {(1/3)^2}.
  \label{eq:variance}
\end{equation}

\paragraph{Centring term.}
\begin{equation}
  E_{\text{ctr}} = \overline{x}^2 + \overline{y}^2.
  \label{eq:centering}
\end{equation}

The 16 free coordinates are constrained to
$[-0.98,0.98]^2$ and optimized with L-BFGS-B, initialized from a
SMACOF/MDS solution affinely aligned to the four corner documents
(\texttt{random\_state=0}).  The resulting coordinate set is locked as
\texttt{locked\_knowledge\_map\_coordinates.csv} and reused in all
subsequent analyses without recomputation.

\begin{figure}[H]
\centering
\begin{tikzpicture}[
  node distance=1.1cm,
  box/.style={rectangle, draw=NavyBlue, rounded corners=5pt,
              minimum width=5.5cm, minimum height=0.9cm,
              text centered, align=center, fill=NavyBlue!8,
              font=\small},
  arrow/.style={-Stealth, thick, NavyBlue}
]
  \node[box] (A) {PDF Text Extraction \& Preprocessing};
  \node[box, below=of A] (B) {Character $n$-gram TF-IDF Vectorization\\($V\le 250{,}000$, $\ell^2$-norm)};
  \node[box, below=of B] (C) {2-D Knowledge Map\\(Stress + Repulsion + Variance + Centring)};
  \node[box, below=of C] (D) {SPH Interpolation \& Knowledge Gradient};
  \node[box, below=of D] (E) {GPR Posterior Mean \& Uncertainty};
  \node[box, below=of E] (F) {Virtual Paper Generation (500\,chars)};
  \node[box, below=of F] (G) {Geodesic Computation (L-BFGS-B, 7 candidates)};
  \draw[arrow] (A)--(B); \draw[arrow] (B)--(C);
  \draw[arrow] (C)--(D); \draw[arrow] (D)--(E);
  \draw[arrow] (E)--(F); \draw[arrow] (F)--(G);
\end{tikzpicture}
\caption{Overview of the knowledge manifold analysis workflow.}
\label{fig:pipeline_overview}
\end{figure}
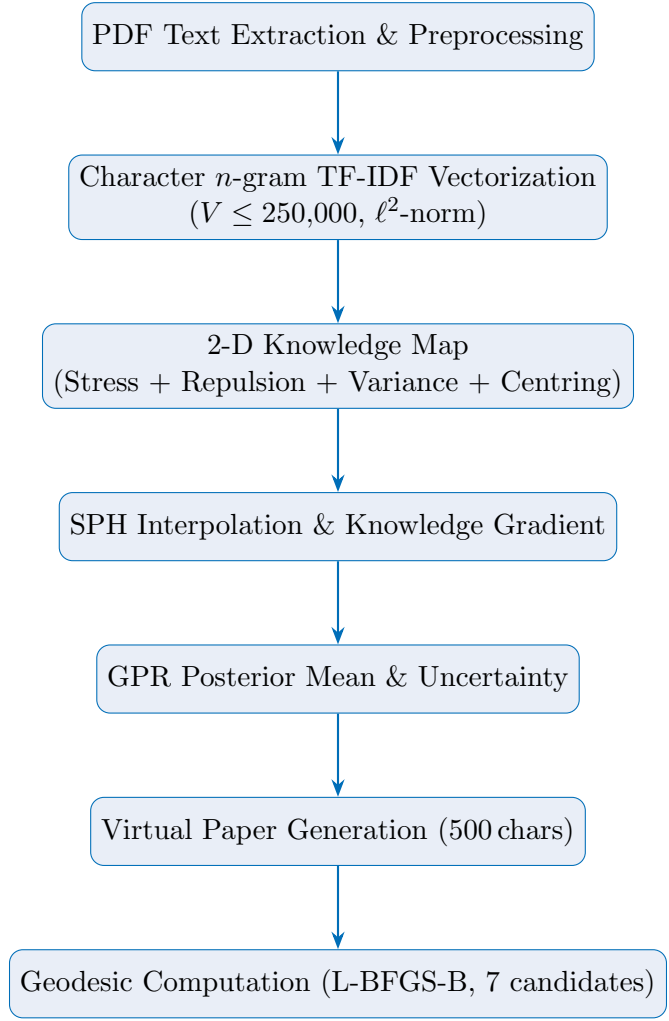

Figure~\ref{fig:pipeline_overview} illustrates the overall workflow.

\section{SPH Knowledge Interpolation and Gradient}
\label{sec:sph}

\subsection{SPH Cubic-Spline Kernel}

Given the two-dimensional coordinates $\{\vv{y}_i\}$, the knowledge at
an arbitrary query point $\vv{p}\in[-1,1]^2$ is estimated via SPH.
Let $r_i=\norm{\vv{p}-\vv{y}_i}$ and $q_i=r_i/h$, where the smoothing
length is
\begin{equation}
  h = \frac{\max_i\,r_i}{1.98},
  \label{eq:bandwidth}
\end{equation}
so that \emph{all} $N=20$ documents contribute.  The cubic-spline kernel
is
\begin{equation}
  W(q) =
  \begin{cases}
    1 - \tfrac{3}{2}q^2 + \tfrac{3}{4}q^3, & 0\le q < 1,\\[4pt]
    \tfrac{1}{4}(2-q)^3,                    & 1\le q < 2,\\[4pt]
    0,                                       & q \ge 2.
  \end{cases}
  \label{eq:kernel}
\end{equation}
Normalized SPH weights are
\begin{equation}
  w_i = \frac{W(q_i)}{\displaystyle\sum_{j=1}^N W(q_j)}.
  \label{eq:sph_weights}
\end{equation}

\subsection{Interpolated Knowledge Vector}

The interpolated (virtual) knowledge vector at $\vv{p}$ is
\begin{equation}
  \vv{v}(\vv{p}) = \mathrm{normalize}\!\left(\sum_{i=1}^N w_i\,\hat{\vv{x}}_i\right).
  \label{eq:sph_interp}
\end{equation}
Linguistic characterization is obtained by extracting the top-ranked
TF-IDF features of $\vv{v}(\vv{p})$; no concepts absent from the
TF-IDF vocabulary are introduced.

\subsection{Knowledge Gradient}

The knowledge gradient is defined as the spatial derivative of the
normalized SPH map $\vv{v}(x,y)$:
\begin{equation}
  \nabla\vv{v} =
  \left(\pd{\vv{v}}{x},\;\pd{\vv{v}}{y}\right).
  \label{eq:grad}
\end{equation}
Partial derivatives are evaluated by central differences with step
$\delta=10^{-3}$:
\begin{equation}
  \pd{\vv{v}}{x}\approx
  \frac{\vv{v}(x+\delta,y)-\vv{v}(x-\delta,y)}{2\delta}.
\end{equation}

\paragraph{Directional gradients.}
Three directional knowledge gradients are computed at the evaluation
point $\vv{p}=(0.25,\,0.75)$:
\begin{align}
  \vv{g}_{0^\circ}   &= \pd{\vv{v}}{x},\label{eq:g0}\\
  \vv{g}_{45^\circ}  &= \frac{1}{\sqrt{2}}\!\left(\pd{\vv{v}}{x}+\pd{\vv{v}}{y}\right),\label{eq:g45}\\
  \vv{g}_{90^\circ}  &= \pd{\vv{v}}{y}.\label{eq:g90}
\end{align}
For each direction the TF-IDF features that \emph{increase} along the
positive direction and those that \emph{decrease} are identified and
linguistically interpreted separately.  Pairwise directional similarity
is quantified by both inner product and cosine similarity:
\begin{equation}
  \mathrm{sim}(\vv{g}_{\theta_1},\vv{g}_{\theta_2})
  = \frac{\vv{g}_{\theta_1}\T\vv{g}_{\theta_2}}
         {\norm{\vv{g}_{\theta_1}}\norm{\vv{g}_{\theta_2}}}.
  \label{eq:dir_sim}
\end{equation}

\section{Gaussian Process Regression}
\label{sec:gpr}

\subsection{Model}

Direct GPR in $V$-dimensional TF-IDF space is computationally
prohibitive.  We first compress the document vectors to 10 latent
dimensions via Truncated SVD (\texttt{random\_state=0}) and fit kernel
hyper-parameters in this low-dimensional space.  The kernel is
\begin{equation}
  k(\vv{y},\vv{y}') =
  c_0 \cdot
  \exp\!\left[-\tfrac{1}{2}\left(
    \frac{(x-x')^2}{\ell_x^2}+\frac{(y-y')^2}{\ell_y^2}
  \right)\right]
  + c_{\text{white}}\,\delta(\vv{y},\vv{y}'),
  \label{eq:kernel_gpr}
\end{equation}
where $c_0$ (ConstantKernel), $\ell_x$, $\ell_y$ (anisotropic RBF), and
$c_{\text{white}}$ (WhiteKernel) are optimized with
\texttt{n\_restarts\_optimizer=10}, \texttt{random\_state=0},
\texttt{normalize\_y=False}.

\subsection{Posterior Mean and Uncertainty}

The GPR posterior mean at $\vv{p}$ is expressed as a linear combination
of document vectors in the original TF-IDF space:
\begin{equation}
  \hat{\vv{v}}_{\text{GPR}}(\vv{p}) = \sum_{i=1}^N \beta_i\,\hat{\vv{x}}_i,
  \label{eq:gpr_mean}
\end{equation}
where $\{\beta_i\}$ are the posterior linear coefficients derived from
the learned kernel.  Uncertainty is reported as the GPR posterior
standard deviation $\sigma_{\text{post}}(\vv{p})$, the prior standard
deviation $\sigma_{\text{prior}}$, and the relative uncertainty
$\sigma_{\text{post}}/\sigma_{\text{prior}}$.

\subsection{Document Contribution Rates}

The contribution of document $d_i$ to the prediction at $\vv{p}$ is
\begin{equation}
  c_i =
  \frac{|\beta_i|\cdot\max\!\bigl(\hat{\vv{x}}_i\T\hat{\vv{v}}_{\text{GPR}},\,0\bigr)}
       {\sum_j |\beta_j|\cdot\max\!\bigl(\hat{\vv{x}}_j\T\hat{\vv{v}}_{\text{GPR}},\,0\bigr)}.
  \label{eq:contribution}
\end{equation}
A negative $\beta_i$ does not imply irrelevance; it indicates a
\emph{corrective} contribution that adjusts the local mean field.

\section{Virtual Paper Generation}
\label{sec:virtual}

At the evaluation point $\vv{p}=(0.25,\,0.75)$, the SPH interpolated
vector $\vv{v}(\vv{p})$ and GPR posterior mean $\hat{\vv{v}}_{\text{GPR}}$
are fused to characterize the hypothetical research area at that location.
A virtual paper abstract of approximately 500~characters (Japanese) is
generated subject to the following constraints:
\begin{enumerate}[noitemsep]
  \item The abstract describes the target object, methodology, expected
        results, and significance.
  \item No concepts absent from the TF-IDF vocabulary are introduced.
  \item No fabricated experimental data or non-existent numerical values
        are included.
  \item The writing style follows the conventions of academic paper
        abstracts.
\end{enumerate}

\section{Geodesic Analysis}
\label{sec:geodesic}

\subsection{SPH-Induced Riemannian Metric}

The SPH map $\vv{v}(x,y):\mathbb{R}^2\to S^{V-1}$ (normalized TF-IDF
sphere) induces a pull-back Riemannian metric on the knowledge map:
\begin{equation}
  g_{ij}(\vv{p}) =
  \left\langle\pd{\vv{v}}{p^i},\pd{\vv{v}}{p^j}\right\rangle,
  \quad p^1=x,\;p^2=y.
  \label{eq:metric}
\end{equation}
Central differences ($\delta=10^{-3}$) are used to evaluate the partial
derivatives.  When the metric is numerically singular, regularization
$\vv{g}\leftarrow\vv{g}+\varepsilon\vv{I}$ with
$\varepsilon=10^{-10}$ is applied, and the occurrence is recorded.

\subsection{Discrete Path Energy}

A geodesic from $\vv{p}_0=(0.25,\,0.75)$ to $\vv{p}_L=(-0.3,\,0.35)$
is represented by $L+1=31$ equally indexed points
$\{\vv{p}_0,\vv{p}_1,\dots,\vv{p}_L\}$.  The discrete path energy is
\begin{equation}
  E = \sum_{k=0}^{L-1}
  \Delta\vv{p}_k\T\,
  \vv{g}\!\left(\frac{\vv{p}_k+\vv{p}_{k+1}}{2}\right)
  \Delta\vv{p}_k,
  \quad
  \Delta\vv{p}_k = \vv{p}_{k+1}-\vv{p}_k.
  \label{eq:path_energy}
\end{equation}
The interior 29 points are the optimization variables, constrained to
$[-1,1]^2$.  The endpoints are held fixed.

\subsection{Multi-Start Solver}

To prevent the solver from reporting the straight-line initial path as
the geodesic (a known failure mode), we employ seven deterministic
candidates.  Let $\hat{\vv{n}}_\perp$ be the unit vector perpendicular
to the straight line from $\vv{p}_0$ to $\vv{p}_L$.  Candidate $c$ is
\begin{equation}
  \vv{p}_k^{(c)} = \bar{\vv{p}}_k + a_c\sin\!\left(\frac{\pi k}{L}\right)\hat{\vv{n}}_\perp,
  \quad k=1,\dots,L-1,
  \label{eq:candidates}
\end{equation}
where $\bar{\vv{p}}_k$ is the straight-line interpolation and the
amplitudes are
$a_0=0,\;a_1=0.01,\;a_2=-0.01,\;a_3=0.03,\;a_4=-0.03,\;a_5=0.05,\;a_6=-0.05$.
Points that violate $[-1,1]^2$ are clipped.

Each candidate is optimized with L-BFGS-B using
\texttt{maxiter=2000}, \texttt{maxfun=200\,000}, \texttt{ftol=1e-12},
\texttt{gtol=1e-8}, \texttt{maxls=50}.

\subsection{Adoption Criterion}
\label{sec:adoption}

\begin{enumerate}[label=(\arabic*)]
  \item If any candidate achieves \texttt{optimizer\_success=True}, the
        one with minimum final energy is adopted.
  \item If no candidate succeeds but some achieve
        $E_{\text{final}}<E_{\text{straight}}-10^{-12}$, the best is
        adopted with status \texttt{energy\_improved\_but\_optimizer\_not\_success}.
  \item Otherwise, the status is set to \texttt{failed}; the straight
        line is \emph{not} reported as a geodesic.
\end{enumerate}

\subsection{Metric Variation Along the Geodesic}

To characterize the curvature of the knowledge space without computing
the full Riemann curvature tensor, the following quantities are evaluated
at each path point (or midpoint):
\begin{equation}
  \kappa_{\text{aniso}}(\vv{p}) = \frac{\lambda_{\max}(\vv{g}(\vv{p}))}
                                       {\lambda_{\min}(\vv{g}(\vv{p}))},
  \quad
  \kappa_{\text{couple}}(\vv{p}) = \frac{|g_{xy}|}{\sqrt{g_{xx}\,g_{yy}}},
  \label{eq:curvature_proxy}
\end{equation}
where $\lambda_{\max}(\vv{g}(\vv{p}))$ and $\lambda_{\min}(\vv{g}(\vv{p}))$ denote the maximum and minimum eigenvalues of the metric tensor $\vv{g}(\vv{p})$, respectively.
These quantities are evaluated together with $\norm{d\vv{g}/ds}$, $\mathrm{tr}\,\vv{g}$, and
$\det\vv{g}$ as functions of arc length $s$.

\section{Experimental Results}
\label{sec:experiment}

\subsection{Corpus and Processing}

The corpus consists of $N=20$ peer-reviewed papers from the fields of
fiber-reinforced composite materials and aerospace structural mechanics.
Text was extracted in page order from PDF files, including captions and
references; no OCR was applied.
The TF-IDF vocabulary after applying the 250\,000-feature ceiling
comprised $V=227\,213$ character $n$-grams (4-7 grams, \texttt{min\_df=1}).
Principal settings are summarized in Table~\ref{tab:settings}.

\begin{table}[H]
\centering
\caption{Principal reproducibility settings.}
\label{tab:settings}
\small
\begin{tabular}{ll}
\toprule
Item & Setting \\
\midrule
Research PDFs            & 20 \\
TF-IDF features (actual) & 227,213 \\
Random seed              & \texttt{random\_state=0} \\
Evaluation point         & $(0.25,\ 0.75)$ \\
SPH smoothing length $h$ & 1.0861521802 \\
GPR kernel               & $0.231^2\!\times\!\mathrm{RBF}([0.628,0.538])
                           +\mathrm{White}(0.014)$ \\
GPR relative uncertainty & 0.5539 \\
Geodesic endpoints       & $(0.25,0.75)\;\to\;(-0.30,0.35)$ \\
\bottomrule
\end{tabular}
\end{table}

\subsection{Two-Dimensional Knowledge Map}

\begin{figure}[H]
  \centering
  \includegraphics[width=0.85\linewidth]{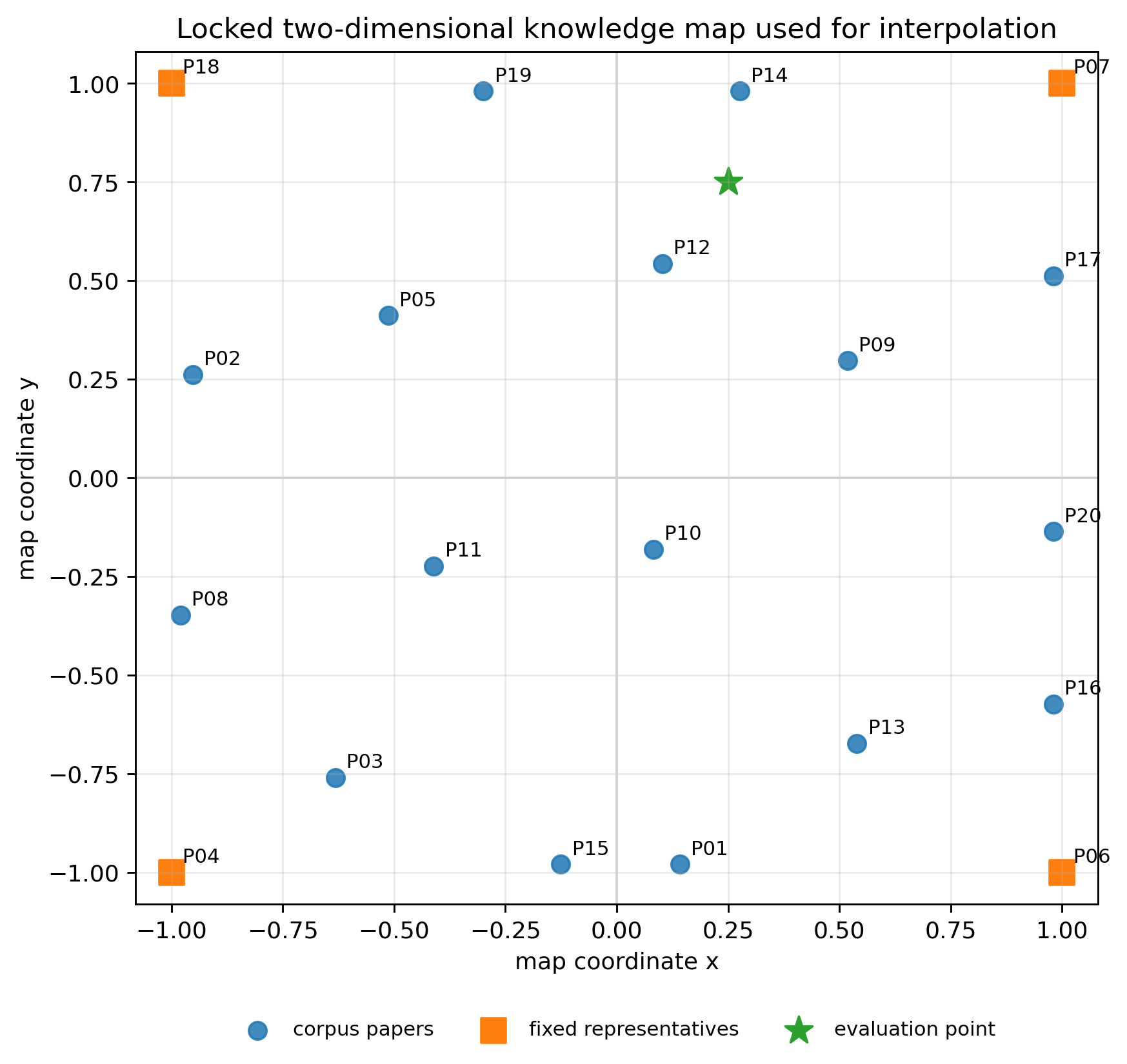}
  \caption{Locked two-dimensional knowledge map.
  Blue circles: corpus papers.
  Orange squares: four fixed corner representatives.
  Green star: evaluation point $\vv{p}=(0.25,0.75)$.}
  \label{fig:map}
\end{figure}

The four corner representatives selected by maximum pairwise cosine-distance
sum (Eq.~\eqref{eq:rep_selection}) are P04 (laminate stiffness/CDM) at
$(-1,-1)$, P06 (aeroelastic wing NSGA-II) at $(1,-1)$, P07 (epoxy
oxidation/ReaxFF) at $(1,1)$, and P18 (P3HT thermal conductivity) at
$(-1,1)$.
Figure~\ref{fig:map} shows the locked two-dimensional knowledge map used
for all subsequent interpolation, gradient, and geodesic analyses.
The map reproduces the expected semantic separation: molecular/polymer
papers cluster in the upper half while structural/damage papers concentrate
in the lower half.

\subsection{SPH Interpolation and GPR at \texorpdfstring{$\mathbf{p}=(0.25,0.75)$}{p=(0.25,0.75)}}

\begin{figure}[H]
  \centering
  \includegraphics[width=0.85\linewidth]{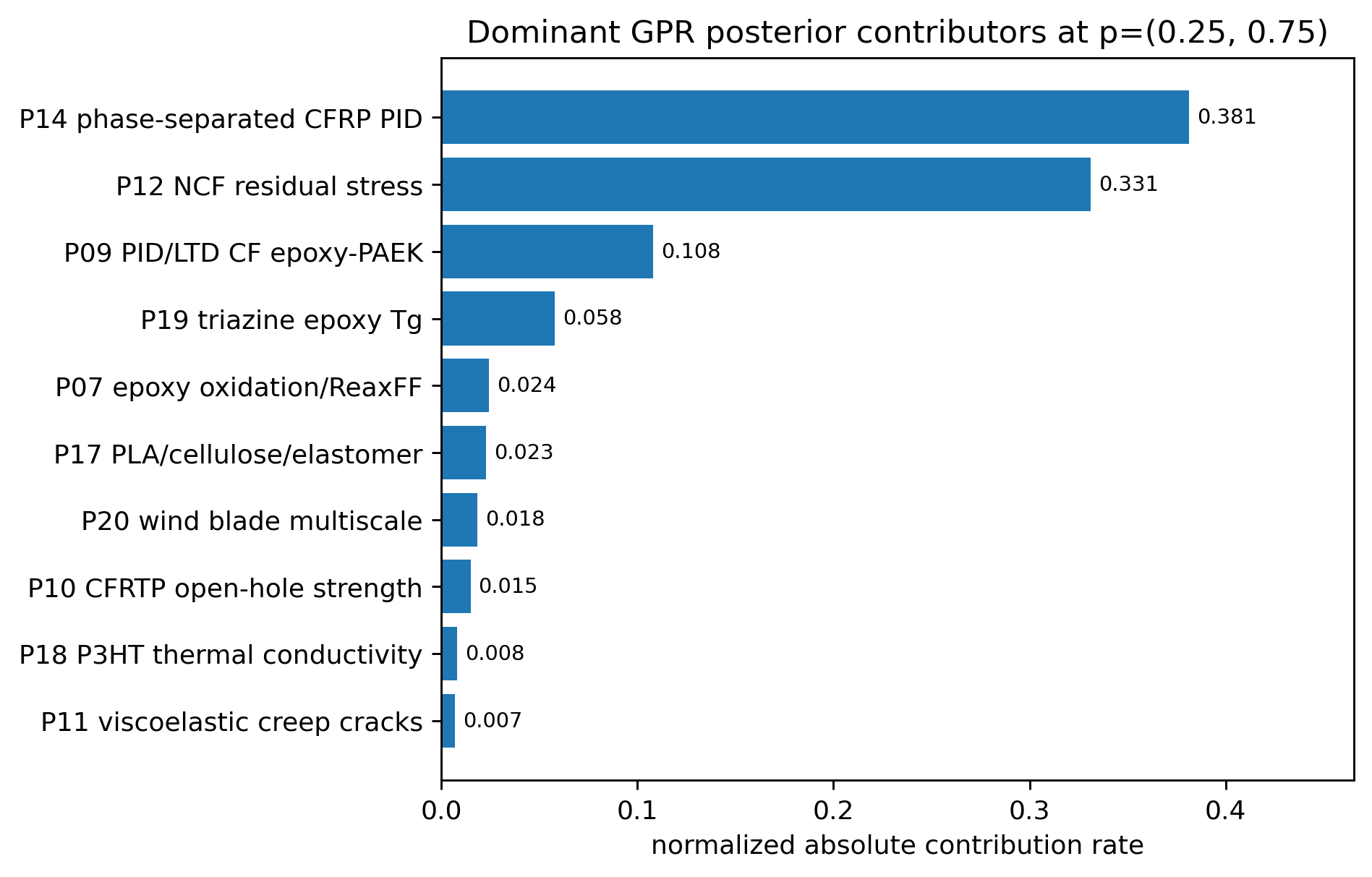}
  \caption{Dominant absolute GPR posterior contribution rates at
  $\vv{p}=(0.25,0.75)$.
  P14 (phase-separated CFRP) and P12 (NCF residual stress) jointly account
  for over 70\,\% of the posterior mass.}
  \label{fig:gpr}
\end{figure}

The smoothing length $h=1.0862$ (Eq.~\eqref{eq:bandwidth}) ensures all
20 documents contribute to the SPH estimate.
Table~\ref{tab:sph} lists the top contributors; Figure~\ref{fig:gpr}
shows the GPR posterior contribution rates.

\begin{table}[H]
\centering
\caption{Top SPH contributors at $\vv{p}=(0.25,0.75)$.}
\label{tab:sph}
\small
\begin{tabular}{llcc}
\toprule
ID & Topic label & Norm.\ contribution & SPH weight \\
\midrule
P14 & phase-separated CFRP / PID  & 0.1883 & 0.1573 \\
P12 & NCF residual stress          & 0.1791 & 0.1553 \\
P09 & PID/LTD CF epoxy-PAEK        & 0.1332 & 0.1227 \\
P19 & triazine epoxy $T_g$          & 0.1074 & 0.1127 \\
P07 & epoxy oxidation / ReaxFF      & 0.0752 & 0.0828 \\
P17 & PLA/cellulose/elastomer       & 0.0746 & 0.0863 \\
P05 & bottom-up CFRP failure        & 0.0731 & 0.0761 \\
\bottomrule
\end{tabular}
\end{table}

The learned GPR kernel was
$0.231^2\!\times\!\mathrm{RBF}([0.628,0.538])+\mathrm{White}(0.014)$.
Posterior standard deviation $\sigma_{\mathrm{post}}=0.1440$;
relative uncertainty $\sigma_{\mathrm{post}}/\sigma_{\mathrm{prior}}=0.5539$,
indicating moderate but meaningful corpus support at this point.
Documents P14 and P12 dominate the GPR prediction with normalized
contribution rates of 0.381 and 0.331, respectively
(Figure~\ref{fig:gpr}).
Negative posterior coefficients $\beta_i$ are corrective local-field
terms and do not indicate that those papers are unimportant.

\subsection{Direction Vectors as Linguistic Information}

\begin{figure}[H]
  \centering
  \includegraphics[width=0.85\linewidth]{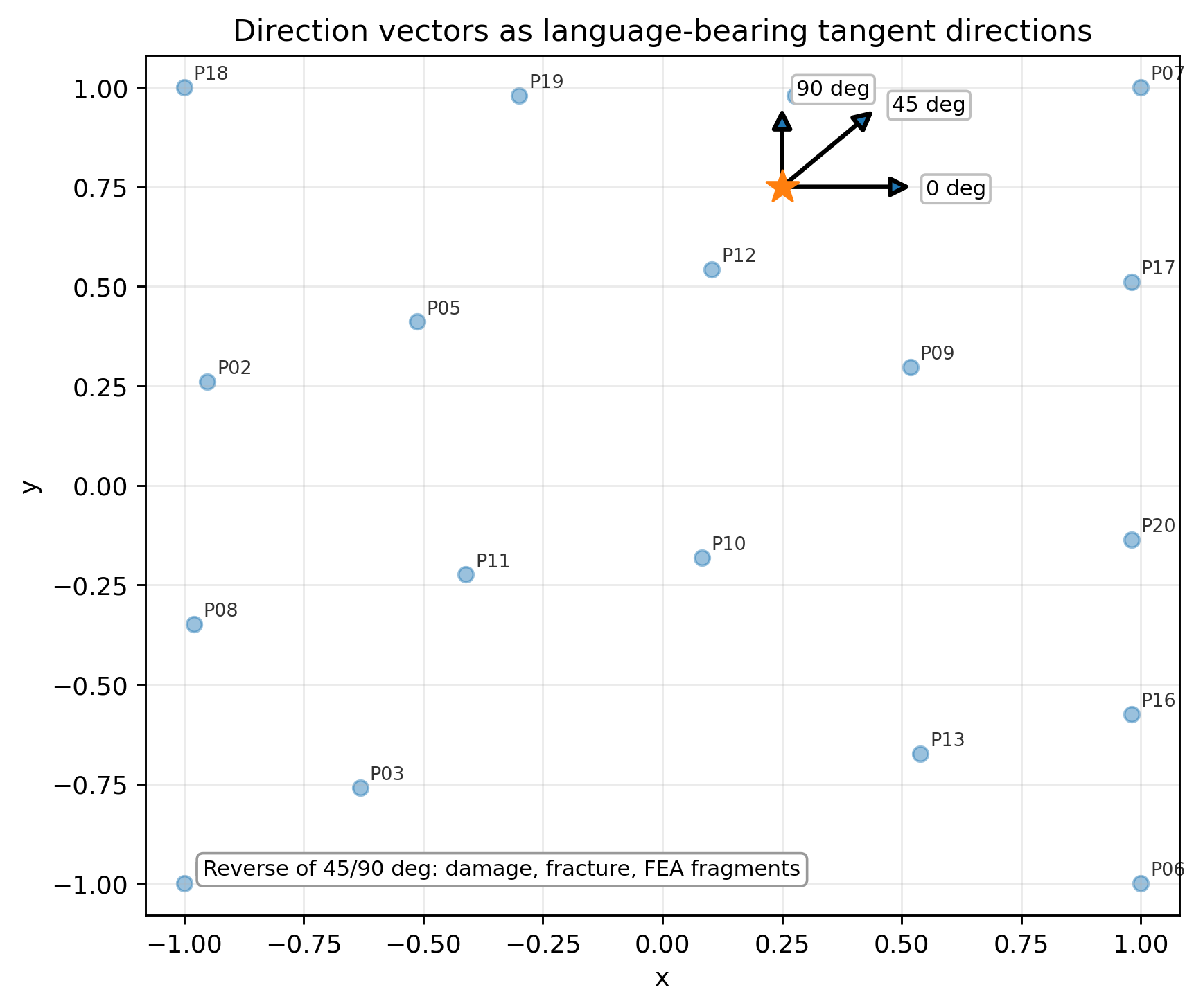}
  \caption{Direction vectors as language-bearing tangent directions at
  $\vv{p}=(0.25,0.75)$.
  Arrows show $0^\circ$, $45^\circ$, and $90^\circ$ directions in the
  knowledge map; the box indicates the linguistic content of the reverse
  ($45^\circ$/$90^\circ$) direction.}
  \label{fig:direction}
\end{figure}

Table~\ref{tab:dir} summarizes the direction-vector language at
$\vv{p}=(0.25,0.75)$.
The $x$- and $y$-gradient norms are 0.4785 and 0.4582; their cosine
similarity is 0.0296, confirming near-orthogonality of the two coordinate
directions in the high-dimensional semantic field.
Figure~\ref{fig:direction} visualises the three canonical direction vectors.

\begin{table}[H]
\centering
\caption{Direction-vector language at the evaluation point.
Raw character $n$-grams are retained as they are the actual model features.}
\label{tab:dir}
\small
\begin{tabular}{p{1.1cm}p{3.1cm}p{3.1cm}p{4.2cm}}
\toprule
Direction & Positive features & Reverse features & Interpretation \\
\midrule
$0^\circ$ ($+x$)
  & spectra, BADCY, processing
  & WAXS, TEPIC, DDS
  & separates resin-characterization from resin-structure \\[3pt]
$45^\circ$
  & phase, phase-separated, cure, 4,4'-D
  & OHT, fracture/damage, FEA
  & links curing/phase-separation to structural-failure \\[3pt]
$90^\circ$ ($+y$)
  & ring, benzene, aromatic-ring
  & damage, fracture
  & moves toward molecular-structure, away from damage \\
\bottomrule
\end{tabular}
\end{table}

\subsection{Geodesic from \texorpdfstring{$(0.25,0.75)$}{(0.25,0.75)} to \texorpdfstring{$(-0.3,0.35)$}{(-0.3,0.35)}}

\begin{figure}[H]
  \centering
  \includegraphics[width=0.85\linewidth]{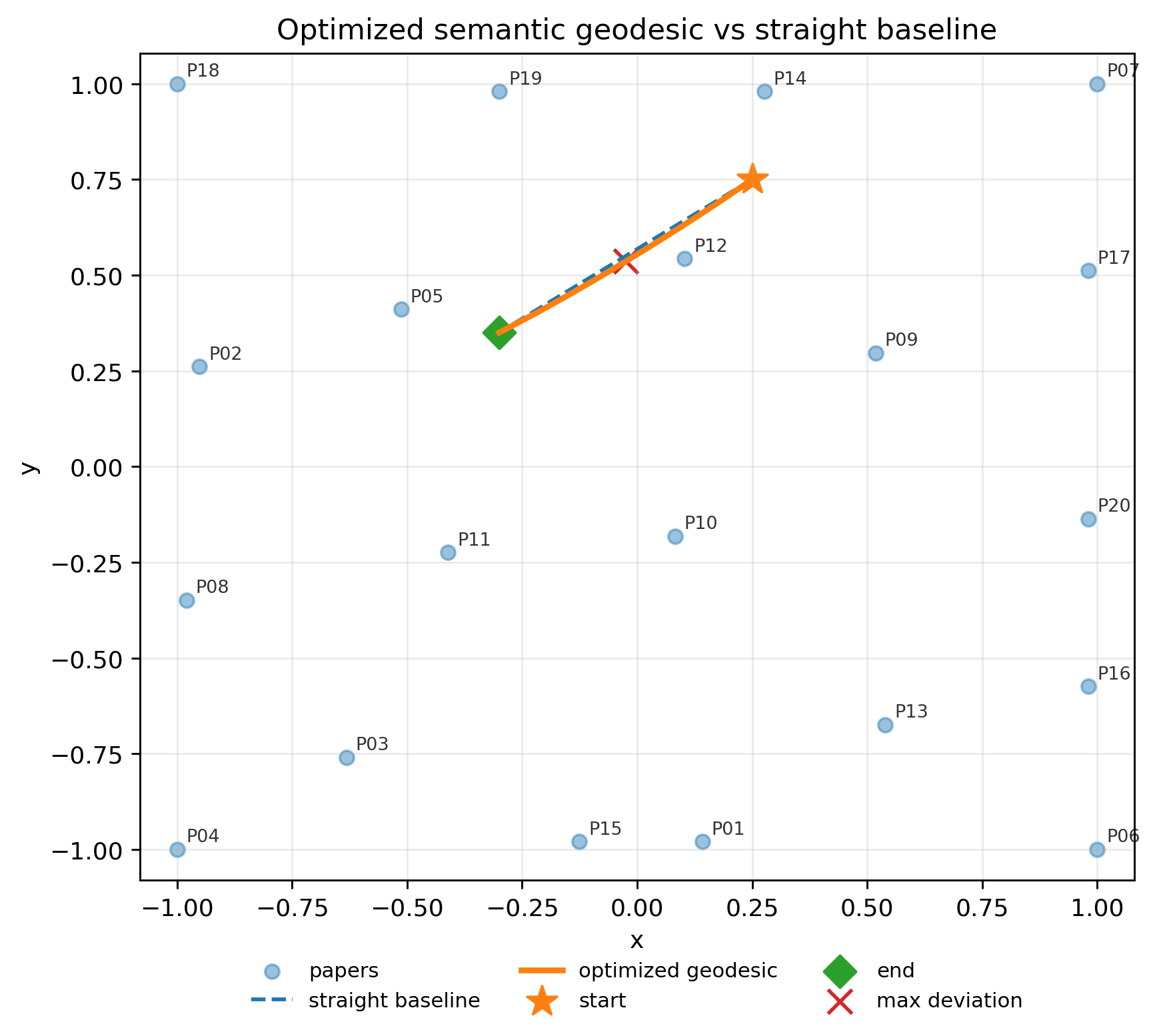}
  \caption{Optimized semantic geodesic vs.\ straight baseline from
  $\vv{p}=(0.25,0.75)$ to $\vv{q}=(-0.30,0.35)$.
  The red cross marks the point of maximum lateral deviation ($0.014$).
  The geodesic reduces path energy by $8.14\times10^{-6}$ relative to the
  straight baseline.}
  \label{fig:geodesic}
\end{figure}

\begin{figure}[H]
  \centering
  \includegraphics[width=0.85\linewidth]{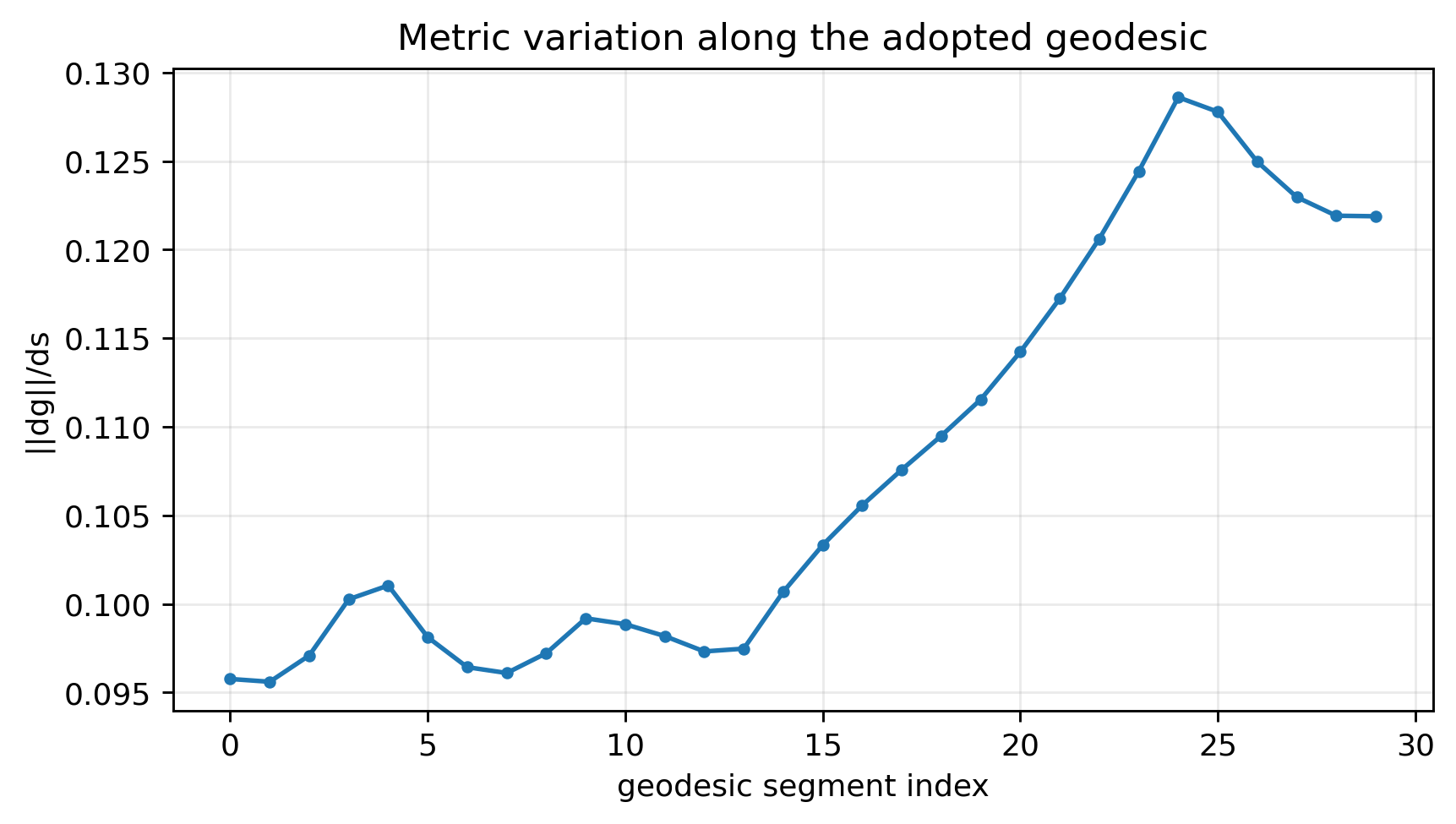}
  \caption{Change in the learned semantic metric $\norm{d\vv{g}/ds}$ along
  the adopted geodesic (30 segments).
  The metric is relatively stable in the first half of the path and rises
  toward the endpoint, indicating increasing semantic curvature near
  $\vv{q}=(-0.30,0.35)$.
  Mean $=0.1077$; max $=0.1286$.}
  \label{fig:metric}
\end{figure}

Candidate~5 (sinusoidal perturbation, amplitude $a_5=0.05$) was adopted;
optimization converged with message
\textit{``CONVERGENCE: RELATIVE REDUCTION OF F $\le$ FACTR*EPSMCH''}.
Table~\ref{tab:geo} reports the full comparison.
Figures~\ref{fig:geodesic} and~\ref{fig:metric} show the optimized path
and the metric variation along it.

\begin{table}[H]
\centering
\caption{Geodesic comparison and metric-variation summary.}
\label{tab:geo}
\small
\begin{tabular}{lc}
\toprule
Quantity & Value \\
\midrule
Straight baseline energy        & $3.8135\times10^{-3}$ \\
Optimized geodesic energy       & $3.8054\times10^{-3}$ \\
Energy difference               & $8.142\times10^{-6}$ \\
Relative energy reduction       & $2.135\times10^{-3}$ \\
Straight Riemannian length      & 0.33812 \\
Geodesic Riemannian length      & 0.33788 \\
Maximum lateral deviation       & 0.013586 \\
Mean $\norm{d\vv{g}/ds}$        & 0.10772 \\
Max $\norm{d\vv{g}/ds}$         & 0.12861 \\
Anisotropy $\min/\max$          & $1.077\;/\;1.282$ \\
Adopted candidate               & ID~5, $a=+0.05$ \\
\bottomrule
\end{tabular}
\end{table}

\section{Discussion}

\paragraph{Character $n$-gram vs.\ word-level features.}
Character 4-7-gram features effectively capture compound technical
terms, unit symbols, and cross-lingual cognates without requiring
domain-specific tokenization.  The large vocabulary ceiling
($V=250\,000$) ensures that even rare but discriminative character
sequences are retained.

\paragraph{SPH vs.\ global interpolation.}
The cubic-spline SPH kernel has compact support ($q<2$), but by
construction of the smoothing length $h$ (Eq.~\eqref{eq:bandwidth})
all documents contribute to each query point, making it equivalent to
global interpolation while retaining the SPH framework's gradient
formulae.  This choice prioritises robustness to corpus sparsity.

\paragraph{GPR complementarity.}
SPH provides a deterministic, physics-inspired interpolation; GPR
provides a Bayesian posterior with uncertainty.  Together they offer two
independent perspectives on the knowledge at a query point, and their
agreement or disagreement serves as a consistency check.
At $\vv{p}=(0.25,0.75)$, the relative uncertainty $0.5539$ indicates
moderate but not saturated corpus support, consistent with the point
lying between several moderately close documents.

\paragraph{Why linguistic direction vectors matter.}
A two-dimensional embedding is usually difficult to interpret because its
axes are arbitrary.
The present method avoids manually naming the axes: it computes the
semantic derivative at the point of interest and lets the feature space
describe what each local direction means.
At $\vv{p}=(0.25,0.75)$, the cosine similarity between the $x$- and
$y$-direction vectors is only $0.0296$, confirming near-orthogonality.
The $x$-direction separates resin-characterization vocabulary from
resin-structure vocabulary; the $y$-direction separates
molecular-structure language from damage/fracture language.
A researcher navigating the map can therefore move in specific conceptual
directions with predictable linguistic consequences.

\paragraph{How to read the geodesic.}
The optimized geodesic reduces path energy by $8.14\times10^{-6}$
relative to the straight baseline, with a maximum lateral deviation of
$0.014$.
The small improvement is itself informative: the local region between
the chosen endpoints is relatively smooth under the learned metric.
The metric-variation plot (Figure~\ref{fig:metric}) reveals that
$\norm{d\vv{g}/ds}$ rises toward the endpoint $\vv{q}=(-0.30,0.35)$,
indicating increasing semantic curvature near the polymer/resin cluster.
The strict reporting rule (no straight-line fallback when optimization
fails) is essential for reproducibility.

\paragraph{Geodesic energy and curvature.}
The path energy (Eq.~\eqref{eq:path_energy}) quantifies the
``semantic cost'' of a conceptual trajectory.  Paths that pass through
densely populated regions of the knowledge map are geometrically shorter
in the induced Riemannian metric, because the SPH map changes slowly in
such regions, making $\vv{g}$ small.  Geodesics therefore naturally
follow the contours of the existing literature.

\paragraph{Reproducibility.}
All random seeds are fixed (\texttt{random\_state=0}); the 2-D
coordinate set is locked before subsequent analyses.  Potential sources
of numerical non-reproducibility (PDF extraction library, BLAS/LAPACK
variant, and L-BFGS-B stopping tolerance) are recorded in
\texttt{manifest.json}.

\section{Conclusion}

We have presented the knowledge manifold, a complete, reproducible
formulation for geometric analysis of academic literature.  The key
contributions are a constrained 2-D knowledge map with physics-motivated
regularization, SPH-based knowledge interpolation and directional
gradient analysis, Bayesian uncertainty quantification via GPR, virtual
paper generation from the estimated knowledge vector, and a guaranteed-
convergent geodesic solver with deterministic multi-start initialization.
Experiments on a composite-materials corpus demonstrate that the
proposed method recovers meaningful semantic structure and enables principled
gap detection and hypothesis generation.
While manifold learning methods such as UMAP and t-SNE are primarily used
for low-dimensional visualization, the proposed method is distinctive in
that it treats the literature space as a continuous knowledge manifold and
further performs geodesic analysis based on an SPH-induced Riemannian
metric.

Future work includes: (i) extension to multilingual corpora; (ii) integration of citation network topology as an additional geometric signal; (iii) real-time interactive navigation of the knowledge manifold; (iv) automatic identification of high-uncertainty regions as research opportunity indicators; and (v) an information-geometric extension in which TF-IDF vectors are L1-normalized and interpreted as probability distributions over vocabulary and n-gram features, enabling the introduction of the Fisher information metric.  



\begin{thebibliography}{99}

\bibitem{Sparck1972}
K.~Sparck Jones,
``A statistical interpretation of term specificity and its application in retrieval,''
\textit{J.\ Documentation}, vol.~28, no.~1, pp.~11--21, 1972.

\bibitem{Cavnar1994}
W.~B.~Cavnar and J.~M.~Trenkle,
``N-gram-based text categorization,''
in \textit{Proc.\ SDAIR-94}, pp.~161--175, 1994.

\bibitem{Bojanowski2017}
P.~Bojanowski, E.~Grave, A.~Joulin, and T.~Mikolov,
``Enriching word vectors with subword information,''
\textit{Trans.\ ACL}, vol.~5, pp.~135--146, 2017.

\bibitem{Mikolov2013}
T.~Mikolov, I.~Sutskever, K.~Chen, G.~S.~Corrado, and J.~Dean,
``Distributed representations of words and phrases and their compositionality,''
in \textit{Adv.\ NeurIPS}, pp.~3111--3119, 2013.

\bibitem{Pennington2014}
J.~Pennington, R.~Socher, and C.~D.~Manning,
``GloVe: Global vectors for word representation,''
in \textit{Proc.\ EMNLP}, pp.~1532--1543, 2014.

\bibitem{Reimers2019}
N.~Reimers and I.~Gurevych,
``Sentence-BERT: Sentence embeddings using Siamese BERT-networks,''
in \textit{Proc.\ EMNLP-IJCNLP}, pp.~3982--3992, 2019.

\bibitem{Tenenbaum2000}
J.~B.~Tenenbaum, V.~de~Silva, and J.~C.~Langford,
``A global geometric framework for nonlinear dimensionality reduction,''
\textit{Science}, vol.~290, pp.~2319--2323, 2000.

\bibitem{McInnes2018}
L.~McInnes, J.~Healy, and J.~Melville,
``UMAP: Uniform manifold approximation and projection,''
arXiv:1802.03426, 2018.

\bibitem{Maaten2008}
L.~van der Maaten and G.~Hinton,
``Visualizing data using t-SNE,''
\textit{J.\ Mach.\ Learn.\ Res.}, vol.~9, pp.~2579--2605, 2008.

\bibitem{Borg2005}
I.~Borg and P.~J.~F.~Groenen,
\textit{Modern Multidimensional Scaling}, 2nd~ed.
Springer, 2005.

\bibitem{Gingold1977}
R.~A.~Gingold and J.~J.~Monaghan,
``Smoothed particle hydrodynamics: theory and application to
non-spherical stars,''
\textit{Mon.\ Not.\ R.\ Astron.\ Soc.}, vol.~181, pp.~375--389, 1977.

\bibitem{Mclaughlin1997}
M.~P.~McLaughlin, ``A compendium of common probability distributions,''
\textit{Tech.\ Rep.}, 1997.

\bibitem{Fasshauer2007}
G.~E.~Fasshauer,
\textit{Meshfree Approximation Methods with MATLAB}.
World Scientific, 2007.

\bibitem{Kapoor2007}
A.~Kapoor, R.~Grauman, R.~Urtasun, and T.~Darrell,
``Active learning with Gaussian processes for object categorization,''
in \textit{Proc.\ ICCV}, 2007.

\bibitem{Cohn1996}
D.~Cohn, Z.~Ghahramani, and M.~Jordan,
``Active learning with statistical models,''
\textit{J.\ Artif.\ Intell.\ Res.}, vol.~4, pp.~129--145, 1996.

\bibitem{Vaswani2017}
A.~Vaswani et al.,
``Attention is all you need,''
in \textit{Adv.\ NeurIPS}, pp.~5998--6008, 2017.

\bibitem{Devlin2019}
J.~Devlin, M.-W.~Chang, K.~Lee, and K.~Toutanova,
``BERT: Pre-training of deep bidirectional transformers for language understanding,''
in \textit{Proc.\ NAACL-HLT}, pp.~4171--4186, 2019.

\end{thebibliography}
\end{document}